\documentclass[
aps,
prl,
reprint,
groupedaddress
]{revtex4-2}

\usepackage{marginnote}

\usepackage{amsmath}
\usepackage{amsfonts}
\usepackage{graphicx}
\usepackage{amssymb}
\usepackage{textcomp}
\usepackage{bm}
\usepackage{dsfont}

\usepackage{pgfplots}

\usepackage{siunitx}

\usepackage{mathtools}

\usetikzlibrary{arrows,shapes,backgrounds,calc,
    positioning,
    intersections}

\definecolor{Blue}{cmyk}{1,0.6,0,0.05}
\definecolor{Red}{cmyk}{0.04,0.87,0.89,0}
\definecolor{Green}{cmyk}{1,0.,1,0}
\definecolor{Yellow}{cmyk}{0,0.2,1,0}
\definecolor{Orange}{cmyk}{0.0,0.69,1,0}
\definecolor{Orange2}{cmyk}{0.0,0.6,1,0.00}
\definecolor{SoftCyan}{cmyk}{0.48,0.,0.07,0.11}

\definecolor{BlenderOrange}{rgb}{1, 0.5, 0.168}
\definecolor{BlenderOrangeDark}{rgb}{1., 0.43, 0.27} 

\definecolor{LaserBlue}{rgb}{0.4,0,1}
\definecolor{LaserMOT}{rgb}{1,0.372,0}
\definecolor{LaserJseven}{rgb}{1,0,0}
\definecolor{LaserJeight}{rgb}{0.35,0,0}

\pgfplotsset{colormap={CM}{color=(white) color=(Blue!50!white) color=(Blue)  color=(Blue!75!black) color=(Blue!50!black)  color=(Blue!25!black) color=(black)}}

\pgfplotsset{colormap={RedToBlue}{color=(Red)  color=(white!50!black) color=(Blue)}}

\pgfplotsset{colormap={CM2}{color=(white) color=(Blue!33!white) color=(Blue!67!white) color=(Blue)  color=(Blue!75!black) color=(Blue!50!black)  color=(Blue!25!black) color=(black)}}
\pgfplotsset{colormap={CM3}{color=(white)  color=(Blue)  color=(black)}}
\pgfplotsset{colormap={CM4}{color=(white)  color=(Red)  color=(black)}}

\DeclareMathOperator{\Tr}{Tr}

\DeclareMathOperator{\atan}{atan}
\DeclareMathOperator{\acos}{acos}
\DeclareMathOperator{\erfc}{erfc}

\DeclareMathOperator{\Imag}{Im}

\newcommand{\fig}[1]{Fig.\,\ref{#1}}

\newcommand{\eqP}[1]{(\ref{#1})}

\newcommand{\bitem}{\begin{itemize}}
\newcommand{\eitem}{\end{itemize}}

\newcommand{\bti}{\begin{tikzpicture}}
\newcommand{\eti}{\end{tikzpicture}}

\newcommand{\ket}[1]{\left| #1 \right>} 
\newcommand{\bra}[1]{\left< #1 \right|} 

\newcommand{\kB}{k_{\text{B}}}

\newcommand{\dd}{\text{d}}

\newcommand{\I}{\text{i}}

\newcommand{\E}{\text{e}}

\newcommand{\pF}{p_{\text{F}}}

\newcommand{\rbold}{\textbf{r}}

\usepackage{xspace}

\newcommand{\mzzero}{m_{z0}}
\newcommand{\pxzero}{p_{x0}}
\newcommand{\pxstar}{p_x^\star}

\newcommand{\EF}{E_{\text{F}}}
\newcommand{\pFm}{p_{\text{F}}^{-}}
\newcommand{\pFp}{p_{\text{F}}^{+}}

\newcommand{\rhoA}{\rho_A}
\newcommand{\px}{p_x}
\newcommand{\vx}{v_x}
\newcommand{\PA}{P_A}
\newcommand{\KA}{K_A}
\newcommand{\psiA}{\psi_A}

\newcommand{\KAvar}{K_A^{\text{var}}}
\newcommand{\HAvar}{H_A^{\text{var}}}
\newcommand{\KAopt}{K_A^{\text{opt}}}

\newcommand{\KABW}{K_A^{\text{BW}}}

\newcommand{\EA}{E_A}

\newcommand{\Ezero}{\mathcal{E}_0}
\newcommand{\EzeroBW}{E_0^{\text{BW}}}

\newcommand{\Omegavar}{\Omega_{\text{var}}}
\newcommand{\bvar}{b_{\text{var}}}
\newcommand{\Qvar}{Q_{\text{var}}}
\newcommand{\Tvar}{T_{\text{var}}}
\newcommand{\muvar}{\mu_{\text{var}}}
\newcommand{\rhovar}{\rho_{\text{var}}}
\newcommand{\Zvar}{Z_{\text{var}}}

\newcommand{\Ezerovar}{E^{\text{var}}_{0}}
\newcommand{\Eonevar}{E_1^{\text{var}}}
\newcommand{\Pimvar}{\Pi_{\text{var}}}
\newcommand{\Pivar}{\Pi_{\text{var}}}

\newcommand{\omc}{\omega_{\text{c}}}
\newcommand{\Nat}{N_{\text{at}}}

\newcommand{\Ac}{A^{\text{c}}}

\newcommand{\Erec}{E_{\text{rec}}}
\newcommand{\vrec}{v_{\text{rec}}}
\newcommand{\omZ}{\omega_{\text{Z}}}

\newcommand{\mstar}{m^\star}

\usepackage[caption=false]{subfig}
\newcommand{\phantomsubfloat}[1]{
    {
        \captionsetup[subfigure]{labelformat=empty}
        \subfloat[][]{#1}
    }%
}
\newcommand{\captionlabel}[1]{\textbf{\protect\subref{#1}},}

\renewcommand{\figurename}{Fig.}

 \makeatletter
 \renewcommand*{\fnum@figure}{{\normalfont\bfseries \figurename~\thefigure}}
 \makeatother

\DeclareSIUnit\gauss{G}

\begin{document}

\reversemarginpar

 \title{
 Realizing the entanglement Hamiltonian of a topological quantum Hall system
 }

 \author{Quentin Redon}
 \thanks{These three authors contributed equally.}
 \author{Qi Liu}
 \thanks{These three authors contributed equally.}
 \author{Jean-Baptiste Bouhiron}
 \thanks{These three authors contributed equally.}
 \author{Nehal Mittal}
 \author{Aurélien Fabre}
\author{Raphael Lopes}
\author{Sylvain Nascimbene}
\email{sylvain.nascimbene@lkb.ens.fr}
 \affiliation{Laboratoire Kastler Brossel,  Coll\`ege de France, CNRS, ENS-PSL University, Sorbonne Universit\'e, 11 Place Marcelin Berthelot, 75005 Paris, France}
 \date{\today}

%
 
 \maketitle
 
\textbf{
Topological quantum many-body systems, such as Hall insulators, are characterized by a hidden order encoded in the entanglement between their constituents. Entanglement entropy, an experimentally accessible single number that globally quantifies entanglement   \cite{vedral_quantifying_1997,eisert_colloquium_2010,islam_measuring_2015}, has been proposed as a first signature of topological order \cite{kitaev_topological_2006,levin_detecting_2006}. Conversely, the full description of entanglement relies on the entanglement Hamiltonian, a more complex object originally introduced  to formulate quantum entanglement in curved spacetime  \cite{nishioka_entanglement_2018,witten_aps_2018}.
As conjectured by Li and Haldane, the entanglement Hamiltonian of a many-body system appears to be directly linked to its boundary properties, making it particularly useful for characterizing topological systems  \cite{li_entanglement_2008}. While the entanglement spectrum is commonly used to identify complex phases arising in numerical simulations \cite{laflorencie_quantum_2016,regnault_entanglement_2017}, its measurement remains an outstanding challenge \cite{dalmonte_entanglement_2022}. Here, we perform a variational approach to realize experimentally, as a genuine Hamiltonian, the entanglement Hamiltonian of a synthetic quantum Hall system \cite{dalmonte_quantum_2018,zache_entanglement_2022}. We use a synthetic dimension \cite{mancini_observation_2015, stuhl_visualizing_2015,chalopin_probing_2020}, encoded in the electronic spin of dysprosium atoms, to implement spatially deformed Hall systems, as suggested by the Bisognano-Wichmann prediction \cite{bisognano_duality_1975,bisognano_duality_1976}. The spectrum of the optimal variational Hamiltonian exhibits a chiral dispersion akin to a topological edge mode, revealing the fundamental link between  entanglement and boundary physics.  Our variational procedure can be easily generalized to interacting many-body systems on various platforms, marking an important step towards the exploration of exotic  quantum systems with long-range correlations, such as fractional Hall states \cite{li_entanglement_2008}, chiral spin liquids \cite{bauer_chiral_2014,he_chiral_2014} and critical systems \cite{calabrese_entanglement_2008}.
}
 
Complex quantum phases of matter, including those exhibiting topological order, are characterized by intricate correlations between their  elementary constituents \cite{amico_entanglement_2008,eisert_colloquium_2010,laflorencie_quantum_2016}. Investigating non-local correlations involves considering a spatial partition  between a subregion $A$ and its complement $\Ac$. The corresponding bipartite entanglement  is described by the properties of the reduced density matrix $\rhoA$, obtained by tracing out particles in $\Ac$.
It is quantified globally by the von Neumann entanglement entropy $S_A=-\text{Tr}\rhoA\log \rhoA$, or the related Renyi entropy \cite{vedral_quantifying_1997,eisert_colloquium_2010}. These entropies  can be measured using various methods, such as  interference  between two copies of a many-body system \cite{daley_measuring_2012,islam_measuring_2015},   randomized measurements \cite{elben_renyi_2018,brydges_probing_2019} and tomography \cite{evrard_observation_2021,tajik_verification_2023}. These techniques have been crucial in investigating the role of entanglement in quantum thermalization \cite{kaufman_quantum_2016}, many-body localization \cite{lukin_probing_2019} and many-body scars \cite{bluvstein_quantum_2022}.

The full description of entanglement goes beyond entanglement entropy and involves a more complex quantity,  the entanglement Hamiltonian $\KA$, defined from $\rhoA\propto\exp(-\KA)$. The concept of the entanglement (or modular) Hamiltonian was first introduced in quantum field theory  \cite{nishioka_entanglement_2018,witten_aps_2018} and has been used to link black hole evaporation and entanglement across its horizon \cite{israel_thermo-field_1976}. It has since been employed  extensively to describe entanglement in strongly correlated quantum systems, especially near phase transitions and at criticality  \cite{calabrese_entanglement_2008}. According to the Li-Haldane conjecture \cite{li_entanglement_2008}, the entanglement Hamiltonian, which characterizes entanglement in the bulk of a physical system, can be brought into direct correspondence with the behavior at its boundary, which is virtually generated by the spatial partition. This connection between  entanglement and edge spectra is a general feature of quantum many-body systems \cite{li_entanglement_2008,poilblanc_entanglement_2010,cirac_entanglement_2011,chandran_bulk-edge_2011,qi_general_2012} and plays a pivotal role in theoretical investigations of topologically ordered phases \cite{regnault_entanglement_2017}, for which  edge physics is ubiquitous.

\begin{figure*}
\includegraphics[scale=0.84,trim=1mm 1mm 0 0]{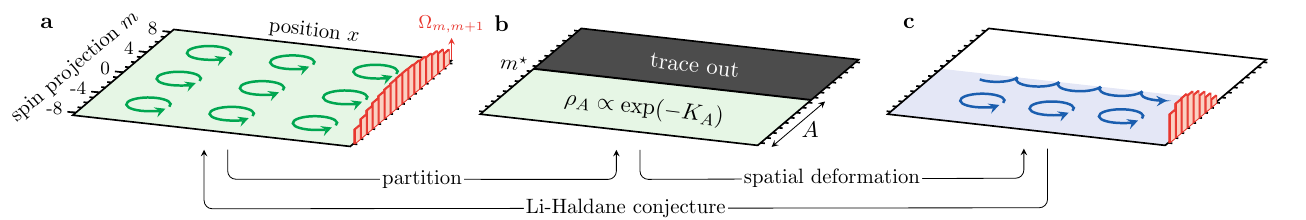}
    \phantomsubfloat{\label{fig:scheme:QH}}
    \phantomsubfloat{\label{fig:scheme:partition}}
    \phantomsubfloat{\label{fig:scheme:KA}}
    \vspace{-2\baselineskip}
\caption{
\textbf{Entanglement Hamiltonian of a synthetic quantum Hall system.} \captionlabel{fig:scheme:QH} Scheme of our system, defined on an $xm$ plane  with a continuous spatial coordinate $x$ and the spin projection $m$ of the electronic spin $J=8$ of dysprosium atoms. A laser-induced spin-orbit coupling, with almost uniform transition amplitudes  $\Omega_{m,m+1}$ between neighboring $m$ states (red bars), gives rise to a quantum Hall effect, illustrated as classical cyclotron orbits.
\captionlabel{fig:scheme:partition} Scheme of the spatial bipartition across the line $m=\mstar=0.5$. The subregion $A$ ($m<\mstar$) is described by a mixed state $\rhoA$, or equivalently the entanglement Hamiltonian $\KA=-\log\rhoA$ (up to a constant).
\captionlabel{fig:scheme:KA} The entanglement Hamiltonian is physically realized as the ground band of a spatially deformed Hall system, which we simulate using spin couplings scaling approximately linearly with the distance $m^\star-m$ near the partition cut (red bars). The atom dynamics, which only takes place in $A$, exhibits a chiral dispersion resembling a virtual edge mode defined by the partition cut, in agreement with the Li-Haldane conjecture. 
\label{fig:scheme}
}
\end{figure*}

Experimental  determination of the entanglement Hamiltonian $\KA$, in particular its spectrum that characterizes the eigenvalues of the reduced density matrix $\rhoA$, has proven to be challenging. Its direct determination by tomography  is limited to small system sizes \cite{choo_measurement_2018}. Other protocols, such as reducing the complexity of tomography with a quasi-locality assumption \cite{kokail_entanglement_2021}, or  interference between many  copies of the same many-body state \cite{pichler_measurement_2016,johri_entanglement_2017,beverland_spectrum_2018}, remain challenging to implement.
An alternative protocol, which we carry out experimentally here, consists in realizing the entanglement Hamiltonian $K_A$ as a genuine Hamiltonian governing the evolution of an auxiliary system, making the entanglement spectrum accessible to  standard spectroscopy techniques \cite{dalmonte_quantum_2018}. The feasibility of this approach is based on the  Bisognano-Wichmann (BW) theorem of quantum field theory, which, for a quasi-local parent Hamiltonian $H$, provides an explicit expression of $\KA$ in terms of a local deformation $\KA=\beta(\textbf{r})H$  \cite{bisognano_duality_1975,bisognano_duality_1976}. In the case of a straight-line bipartition, the deformation factor $\beta(\textbf{r})$ is given by the distance between a generic point $\textbf{r}$ and the partition cut.  Although the BW theorem only applies strictly to continuous systems, it  still provides an excellent approximation of $\KA$ for lattice systems \cite{giudici_entanglement_2018}, making the physical realization of entanglement Hamiltonians accessible to locally tunable quantum simulators.

In this work, we investigate the entanglement properties of a quantum Hall system using an ultracold gas of dysprosium atoms (bosonic isotope $^{162}$Dy). We use the large spin $J=8$ of this magnetic atom to encode a synthetic dimension defined by its projection $m$, with $m$ integer, $|m|\leq J$ (\fig{fig:scheme:QH}). We use a laser-induced effective magnetic field, acting in the $xm$ plane, to simulate a quantum Hall effect \cite{mancini_observation_2015, stuhl_visualizing_2015,chalopin_probing_2020}, and explore spatial entanglement properties for a bipartition between a domain $A$, defined as $m<\mstar=0.5$, and its complement $\Ac$ (\fig{fig:scheme:partition}). Our approach consists of three steps. (i) We experimentally characterize the entanglement Hamiltonian $\KA$ of the system (both its eigenspectrum and eigenstates), by inferring it from the properties of single-particle states. (ii) We implement a family of spatially deformed Hall systems described by Hamiltonians $\HAvar$ which operate only in subregion $A$, and use a variational approach to realize, among the $\HAvar$'s, the optimal approximation of  $\KA$. (iii) We measure the energy spectrum of the optimal $\HAvar$ and find a chiral dispersion resembling a topologically protected edge mode   (\fig{fig:scheme:KA}).

\begin{figure}
\includegraphics[scale=0.8,trim=3.5mm 1mm 0 0]{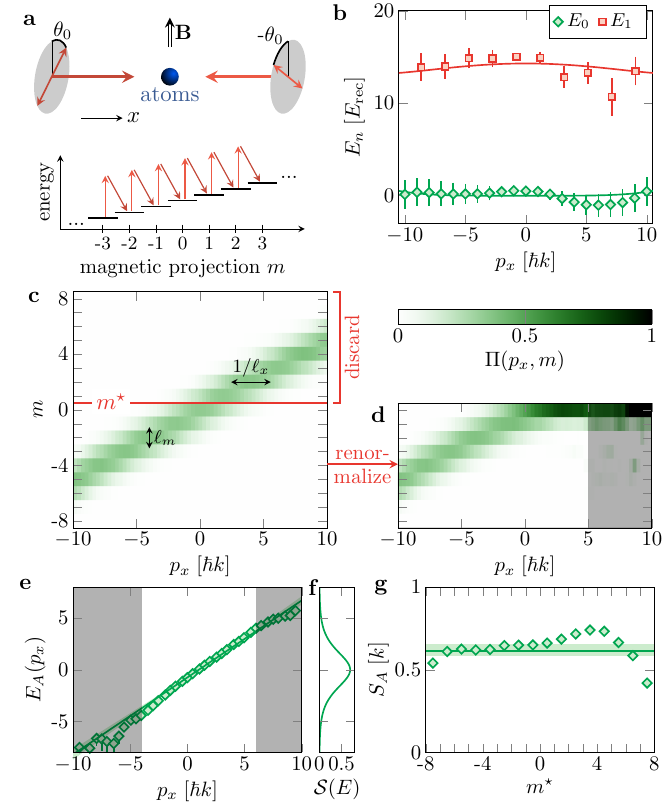}
    \phantomsubfloat{\label{fig:QH:scheme}}
    \phantomsubfloat{\label{fig:QH:dispersion_relation}}
    \phantomsubfloat{\label{fig:QH:projection_probability}}
    \phantomsubfloat{\label{fig:QH:projection_probability_A}}
    \phantomsubfloat{\label{fig:QH:EA}}
    \phantomsubfloat{\label{fig:QH:SE}}
    \phantomsubfloat{\label{fig:QH:SA}}
    \vspace{-2\baselineskip}
\caption{
\textbf{Synthetic quantum Hall system.}
\captionlabel{fig:QH:scheme} Scheme of the laser configuration, with a pair of counterpropagating lasers inducing two-photon spin transitions. The orientation of the linear polarizations ensures quasi-uniform transition amplitudes.
\captionlabel{fig:QH:dispersion_relation} Dispersion relation of our synthetic quantum Hall system. The ground-band energy is measured by integrating the mean velocity $\langle\vx\rangle$ (green diamonds). The first excited band is obtained by measuring the frequency of the cyclotron oscillation subsequent to a weak velocity kick (red squares). The lines show the theoretical bandstructure computed for a coupling $\Omega=3.6\,\Erec/\hbar$.
\captionlabel{fig:QH:projection_probability}\captionlabel{fig:QH:projection_probability_A} Spin projection probabilities $\Pi(\px,m)$ (resp. $\Pi_A(\px,m)$) for the state $\ket{\psi(\px)}$ (resp. its reduction $\ket{\psi_A(\px)}$ to subregion $A$). The widths of the distribution $\Pi(\px,m)$ along $m$ and $\px$ give access to the magnetic lengths $\ell_m$ and $\ell_x$.
\captionlabel{fig:QH:EA} Pseudo-spectrum $\EA(\px)$ derived from the spin projection probabilities $\Pi(\px,m)$. The green solid line is the linear variation given in \eqP{eq:EA}. In d,e, the gray areas indicate the range of momenta for which the state $\ket{\psi_A}$ or pseudo-energy $\EA$ cannot be measured reliably.
\captionlabel{fig:QH:SE} Contribution $\mathcal{S(E)}$ to the entanglement entropy from a mode of energy $E$.
\captionlabel{fig:QH:SA} Entropy per unit length inferred from the pseudo-spectrum, for different positions of the partition cut $\mstar$. The solid line is the expected entropy for our system. Error bars represent the 1-$\sigma$ statistical error.
\label{fig:QH}
}
\end{figure}

\medskip
\noindent{\textbf{Realization of an atomic quantum Hall system}}\\
\noindent The first step involves generating a synthetic quantum Hall system and characterizing its entanglement Hamiltonian $\KA$.
The Hall system is produced using a laser configuration shown in \fig{fig:QH:scheme} \cite{chalopin_probing_2020}. A two-photon process, involving a pair of lasers  counterpropagating along the $x$-axis, induces a spin transition $m\rightarrow m+1$ together with an  $x$-velocity kick $-2\vrec$, so that the canonical momentum $\px=M(\vx+2\vrec m)$ is conserved. Here, $\vrec=\hbar k/M$ is the recoil velocity, $M$ is the atomic mass, and $k=2\pi/\lambda$ is the photon momentum for a wavelength $\lambda=\SI{626.1}{\nano\meter}$.  The atom dynamics are governed by the Hamiltonian
\begin{equation}\label{eq:H_QH}
H=\frac{(\px-2\hbar k J_z)^2}{2M}-\hbar\Omega J_x+Q J_z^2,
\end{equation}
where the quadratic Zeeman field $Q=-\hbar\Omega/(2J+3)$ is optimized to flatten the ground energy band (Methods). Considering the spin projection $m$ as a synthetic dimension, the dynamics in the $xm$ plane map to those of a particle of charge $q=1$ and subjected to a magnetic field $B=2\hbar k$, where the coupling $J_x$ plays the role of kinetic energy along $m$. The single-particle spectrum is organized into quasi-flat energy bands, similar to Landau levels. (\fig{fig:QH:dispersion_relation}).

To analyze the ground-band properties, we prepare  thermal gases with a narrow momentum distribution centered around a mean value $\px$, diluted sufficiently to ensure  negligible interactions between atoms.
We measure their spin-resolved velocity distribution by imaging the gas after free expansion  in the presence of a magnetic gradient, providing the spin projection probabilities $\Pi(\px,m)$ and the mean velocity $\langle \vx\rangle(\px)$. By integrating the velocity, we reconstruct the ground-band energy as $E_0(\px)=\int\dd\px\,\langle \vx\rangle$. We also measure the gap $\omc(\px)$ to the first excited band by monitoring cyclotron center-of-mass dynamics after a quench of small amplitude, hence the first excited band $E_1=E_0+\hbar\omc$ (Methods).

We show in \fig{fig:QH:dispersion_relation} the reconstructed bandstructure for a coupling $\Omega=3.6(1)\,\Erec/\hbar$, where $\Erec=\tfrac{1}{2}M\vrec^2$ is the recoil energy. We focus  on the bulk mode region $|\px|<10\hbar k$, where the population of extremal states $m=\pm J$ is negligible. We observe quasi-flat energy bands that are reminiscent of Landau levels, with a mean cyclotron gap of $\omc=14.2(6)\Erec/\hbar$. Additionally, we  measure the spin projection probabilities $\Pi(\px,m)$ (\fig{fig:QH:projection_probability}) and observe localized distributions along $m$, with a characteristic width $\ell_m=1.53(5)$, which corresponds to the magnetic length along $m$. Similar to Landau orbitals in continuous Hall systems, the center of these distributions varies approximately linearly with $\px$, with a characteristic momentum width inversely proportional to the magnetic length $\ell_x=0.33(2)/k$.

\medskip
\noindent{\mbox{\textbf{Entanglement Hamiltonian characterization}}}\\
\noindent The entanglement Hamiltonian $\KA$ of a quantum Hall insulator can be deduced from the properties of its single-particle orbitals \cite{fidkowski_entanglement_2010,dubail_real-space_2012}. We justify this connection for a generic Hall system whose  bandstructure is indexed by a momentum $\px$, which applies both to Landau levels and our experimental system. It relies on (i) the simplified structure of a fermionic band insulator, which can be expressed as a product state $\bigotimes_{\px}\ket{1\!:\!\psi(\px)}$, where $\px$ indexes the states of the Fermi sea and $\ket{n\!:\!\psi({\px})}$ has $n$ particle in the orbital $\ket{\psi({\px})}$ ($n=0,1$), and (ii) the fact that the partition cut, defined by the line $m=\mstar$, preserves the $x$-translational symmetry and thus the conservation of momentum ${\px}$. Consequently,  the reduced density matrix   retains a factorized form
\begin{align*}\label{eq:rhoA}
 \rhoA=\bigotimes_{\px}\big[&(1-\PA(\px))\ket{0:\psi_A(\px)}\bra{0:\psi_A(\px)}\\[-1mm]
 &\;+
 \PA(\px) \ket{1:\psi_A(\px)}\bra{1:\psi_A(\px)}\big],
\end{align*}
where $\PA({\px})=\int_A\dd\mathrm\rbold |\langle\rbold\ket{\psi({\px})}|^2$ is the probability for a particle in the state $\ket{\psi(\px)}$ to be located in $A$. The state $\ket{\psi_A({\px})}$ is obtained by restricting $\ket{\psi({\px})}$  to $A$ and renormalizing, i.e.
\begin{align*}
\langle\rbold\ket{\psi_A({\px})}&=\langle\rbold\ket{\psi({\px})}/\sqrt{\PA(\px)}\quad\text{for}\quad\rbold\in A,\\
&=0\quad\text{for}\quad\rbold\in A^c.
\end{align*} 
Alternatively, the reduced system can be characterized by the entanglement Hamiltonian $\KA=-\log(\rhoA)$  (up to a constant), as
\begin{equation}
 \label{eq:KA}
 \KA=\sum_{\px}\EA({\px}) n(\px),\quad \EA({\px})=\log\frac{1-\PA({\px})}{\PA({\px})},
\end{equation}
where $n(\px)$ is the occupation number of $\ket{\psi_A({\px})}$.
The dispersion relation $\EA(\px)$ is the so-called single-particle pseudo-energy spectrum of $\KA$ \cite{dubail_real-space_2012}.

We employ the expressions outlined above to determine the properties of the entanglement Hamiltonian of our experimental system, namely its spectrum and eigenstates. The measured spin projection probabilities $\Pi(\px,m)$  yield the probability $\PA(\px)$, and hence the pseudo-energy $\EA(\px)$ (\fig{fig:QH:EA}). It is worth noting that access to $\EA$ requires accurate discrimination between $\PA$ and the values of zero and one, which is feasible for pseudo-energies $|\EA|<5$, i.e. when $\PA$ and $1-\PA$ are at least $1\%$. 
Over this range, the measured pseudo-energy increases  approximately linearly with momentum, consistent with the expected chiral dispersion \cite{oblak_equipartition_2022}
\begin{equation}\label{eq:EA}
 \EA(\px)\simeq\frac{4}{\pi}(\px-\pxstar) \ell_x,
\end{equation}
where $\pxstar=2 \hbar k\mstar=\hbar k$.
Note that for $|\EA|\gg 1$, deviations from this linear variation are expected, but we cannot resolve them given our signal-to-noise ratio.

We also characterize the eigenstates $\ket{\psi_A({\px})}$ through their spin projection probabilities $\Pi_A(\px,m)$ 
(\fig{fig:QH:projection_probability_A}). Note that the probabilities do not provide the complex phase of ground-band wavefunctions, but the latter can be assumed real positive since the Hamiltonian \eqP{eq:H_QH} is real \cite{ribeiro_exact_2008}.  While for negative momenta the probabilities $\Pi_A(\px,m)$ and $\Pi(\px,m)$ almost coincide,  for positive momenta the pseudo-eigenstates $\ket{\psi_A(\px)}$ are found to be localized at the edge $m=0$ of subregion $A$.

The observed behaviors of both $\EA(\px)$ and $\ket{\psi_A(\px)}$ are reminiscent of that of a ballistic edge mode, which  would occur in the presence of a true edge  located at $\mstar$. This finding illustrates the Li-Haldane conjecture, which asserts that the entanglement spectrum reflects the boundary physics, through the virtual edge induced by the partition \cite{li_entanglement_2008,fidkowski_entanglement_2010,dubail_real-space_2012}.

Besides its chiral dispersion, the pseudo-spectrum $\EA(\px)$ can be used to determine the entanglement entropy. The latter corresponds to the thermodynamic entropy of a thermal state of $\KA$, with a  temperature $T=1$ and a chemical potential $\mu=0$. A single-particle orbital of energy $E$ contributes to the entropy as $\mathcal{S}(E)=\mathcal{H}(\langle n\rangle)$, with $\mathcal{H}(u)=-u\log u-(1-u)\log(1-u)$ and  the average occupation $\langle n\rangle=1/(1+\E^{E})$. The function $\mathcal{S}(E)$ is maximum at $E=0$, i.e. for a state equally distributed between $A$ and $A^c$ (\fig{fig:QH:SE}). As the momentum density of states is proportional to the length  $L_x$ of the system along $x$, summing all momentum contributions yields an entropy that scales exactly as $L_x$. The measured pseudo-spectrum $\EA(\px)$ yields an entropy per unit length $S_A=0.66(3)k$ (Methods), which is consistent with the law $S_A\simeq0.203/\ell_x=0.63(3)k$ expected for a continuous Hall system \cite{rodriguez_entanglement_2009}. We repeated this analysis for different positions of the partition cut $\mstar$, which changes the `volume' of subregion $A$  (\fig{fig:QH:SA}). We observed only a weak variation of $S_A$ with $\mstar$, which is consistent with an entropy scaling with the length $L_x$ of the boundary only and corresponds to the `area' law of short-range entangled systems \cite{eisert_colloquium_2010,tajik_verification_2023}.

\medskip
\noindent{\mbox{\textbf{Variational realization of $\KA$}}}\\
\noindent With the entanglement Hamiltonian now determined, our next step is to create a physical Hamiltonian that can match it. To this end, we employ a variational approach, inspired by \cite{kokail_quantum_2021,zache_entanglement_2022}, and analyze a family of Hamiltonians $\HAvar$ that are defined on the subregion $A$.

The choice of the variational Hamiltonian  is guided by the BW theorem  \cite{bisognano_duality_1975,bisognano_duality_1976}, which provides an explicit expression of $\KA$ as a symmetrized product
\begin{equation}
\KABW = \frac{4}{\sqrt{\pi}}\left\{\frac{\mstar-J_z}{\ell_m}, \frac{H-\Ezero}{\hbar\omc}\right\}\label{eq:KABW}
\end{equation}
between the original Hamiltonian $H$ and a local deformation factor, which equals the distance $\mstar-m$ between a generic point $(x,m)$ and the partition line at $\mstar$. The choice of the energy offset $\Ezero$ is discussed in Methods.  For a continuous quantum Hall system described by Landau levels, the ground band of such a BW Hamiltonian coincides at low energy with the entanglement Hamiltonian $\KA$. In our system with a discrete synthetic dimension, we expect that the ground band of $\KABW$ only provides an approximation of $\KA$ (Methods).
Note that spin transitions governed by $\KABW$ are solely described by the operator $\{m^\star-J_z,J_x\}=(J_+J_z+J_zJ_-)/2$, which does not couple the states $m=0$ and 1. As a result, the Hamiltonian $\KABW$ acts separately on the two subregions $A$ and $A^c$, and we consider in the following  its restriction to $A$.

\begin{figure}
\includegraphics[scale=0.86,trim=1mm 0 0 0]{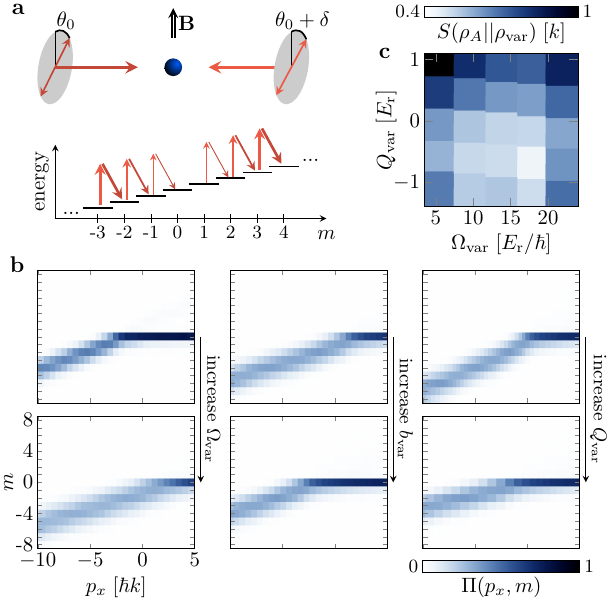}
    \phantomsubfloat{\label{fig:variational:scheme}}
    \phantomsubfloat{\label{fig:variational:parameters}}
    \phantomsubfloat{\label{fig:variational:entropy}}
    \vspace{-2\baselineskip}
\caption{
\textbf{Optimization of the variational Hamiltonian.}
 \captionlabel{fig:variational:scheme} Scheme of the laser configuration used to realize the variational Hamiltonian $\HAvar$. The coupling between the magnetic states $m=0$ and 1 cancels for a polarization mismatch $\delta\simeq\SI{3}{\degree}$.
 \captionlabel{fig:variational:parameters} Spin projection probabilities measured for different choices of the variational parameters $(\Omegavar,\bvar,\Qvar)$, expressed in units of $\Erec$: left column, $\bvar=0.0(3)$, $\Qvar=-0.1(2)$, $\Omegavar=5.6(3)$ and $21.4(10)$; middle column, $\Omegavar=13.9(8)$, $\Qvar=-0.1(2)$, $\bvar=-8.0(3)$ and $8.0(3)$; right column, $\Omegavar=13.9(8)$, $\bvar=0.0(3)$,  $\Qvar=-2.1(2)$ and $1.9(2)$.
 \captionlabel{fig:variational:entropy} Relative entropy as a function of $\Omegavar$ and $\Qvar$. The other variational parameters $\bvar$, $\Tvar$ and $\muvar$ are optimized for each  point. 
\label{fig:variational}
}
\end{figure}

Realizing $\KABW$ in physical systems requires the implementation of operators of third order, such as $p_x^2 J_z$, $p_x J_z^2$, and $J_z^3$, which can be challenging with the standard atom-light interaction toolbox. Instead, we implement a family of deformed Hall systems described by the Hamiltonian
\begin{equation}\label{eq:HAvar}
\HAvar\!=\!\frac{(\px\!-\!2\hbar k J_z)^2}{2M}-\frac{\hbar\Omegavar}{2J}(J_+J_z+\text{h.c.})+\bvar J_z+\Qvar J_z^2,
\end{equation}
parametrized by the spin-hopping amplitude $\Omegavar$ and the linear and quadratic Zeeman fields $\bvar$ and $\Qvar$.
We use rank-2 tensor light shifts to generate the second-order spin coupling $(J_+J_z+J_zJ_-)$,  based on the  laser configuration shown in  \fig{fig:variational:scheme} (Methods).

Following a similar procedure as with the undeformed Hall system, we measure the ground and first excited band properties of $\HAvar$ for different choices of the variational parameters. When ramping up the momentum $\px$ across the ground band, the states $m>0$ show negligible projection probabilities, confirming the decoupling between $A$ and $A^c$ (Methods).
We show in \fig{fig:variational:parameters} the qualitative effect of the variational parameters $\Omegavar$, $\bvar$ and $\Qvar$, which  respectively control the width along $m$, a momentum shift and a momentum scaling  of the ground-band distributions $\Pivar(\px,m)$. We optimize these parameters by minimizing the relative entropy $S(\rho_A||\rhovar)\equiv\Tr[\rho_A(\log\rho_A-\log\rhovar)]$ between  the target reduced density matrix $\rho_A=\exp(-\KA)/Z_A$ and a thermal density matrix
\begin{equation}
\rhovar\equiv\frac{1}{\Zvar}\exp\left(-\KAvar\right),\quad\KAvar=\frac{\HAvar- \muvar}{\kB\Tvar},\label{eq:rhovar}
\end{equation}
where we introduce two additional variational parameters, the temperature $\Tvar$ and the chemical potential $\muvar$. Physically, the relative entropy quantifies the loss of information when approximating  $\rho_A$ with $\rhovar$ \cite{vedral_role_2002}. It can be computed from the measured  spin projection probabilities $\Pimvar(\px,m)$ of the ground band, as well as the measured dispersion relations $\Ezerovar(\px)$ and $\Eonevar(\px)$ of the ground and first excited bands (Methods). The relative entropy is minimized when, simultaneously, (i) the projection probabilities $\Pimvar(\px,m)$ and $\Pi_A(\px,m)$ coincide (ii) the ground band dispersion $\Ezerovar(\px)$ matches the pseudo-spectrum $\EA(\px)$  (iii) the thermal excitation of excited bands is negligible, which occurs when $\Eonevar(\px)\gg1$.

\begin{table}
\begin{tabular}{ccccc}
\hline
\hline
&$\hbar\Omegavar$&$\bvar$&$\Qvar$&$\kB\Tvar$\\
\hline
optimal $\KAvar$ &17.6(11)&-10.8(3)&-0.7(2)&3.7(1)\\
quadratic approx. $\KABW$&19.3&-10.6&-0.47&3.1\\
\hline
\hline
\end{tabular}
\caption{
\textbf{Optimal variational parameters.}
Comparison between the optimal variational parameters and those of a quadratic approximation of the BW Hamiltonian. The parameters are given in units of $\Erec$. Since the experimental bandstructure is obtained up to an arbitrary energy offset, the absolute value of the chemical potential $\muvar$ is irrelevant.
\label{tab:optimal}
}
\end{table}

The relative entropy, measured for a range of variational parameters, exhibits a minimum for the parameters listed in Tab.\,\ref{tab:optimal} (see  its variation with $\Omegavar$ and $\Qvar$ in \fig{fig:variational:entropy}). To test the link between our optimal variational Hamiltonian and the BW ansatz $\KABW$, we expand  the latter to second order in the operators $\px$ and  $J_z$, so that it belongs to the family of $\HAvar$'s (Methods). The corresponding variational parameters are in good agreement with the optimal ones determined experimentally  (Tab.\,\ref{tab:optimal}).

We present in Methods an alternative comparison with $\KABW$, by fitting the factors $\beta(m)$ so that the deformed Hamiltonian $\{\beta(J_z),H\}$ best matches our optimal variational Hamiltonian. We find that, for magnetic projections $m$ near the partition cut $\mstar$, the fitted factors $\beta(m)$ agree with the linear behavior $\propto(\mstar-m)$ expected from the BW ansatz.

\begin{figure}
\includegraphics[scale=0.9,trim=1mm 2mm 0mm 1mm]{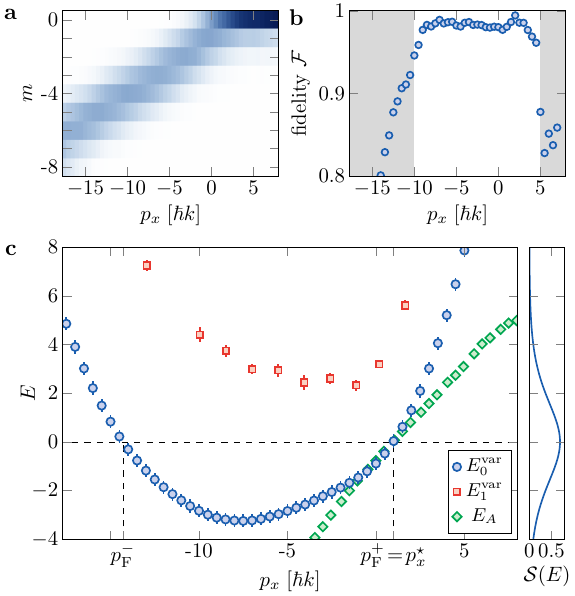}
    \phantomsubfloat{\label{fig:optimum:pops}}
    \phantomsubfloat{\label{fig:optimum:infidelity}}
    \phantomsubfloat{\label{fig:optimum:energy}}
    \vspace{-2\baselineskip}
\caption{
\textbf{Properties of the optimal Hamiltonian.}
 \captionlabel{fig:optimum:pops} Spin projection probabilities $\Pimvar(\px,m)$ of the optimal variational Hamiltonian.
 \captionlabel{fig:optimum:infidelity} Fidelity between the latter and the target probabilities $\Pi_A(\px,m)$. The gray areas indicate momenta outside the interval over which the optimization is performed.
  \captionlabel{fig:optimum:energy} Dispersion relation of the optimal Hamiltonian, measured for the ground and first excited bands (blue disks and red squares, respectively). Green diamonds show the pseudo-spectrum already displayed in \fig{fig:QH:EA}.
  Error bars represent the 1-$\sigma$ statistical error.
\label{fig:optimum}
}
\end{figure}

\medskip
\noindent{\textbf{Optimum variational Hamiltonian}}\\
\noindent The properties of the  optimum  Hamiltonian are shown in \fig{fig:optimum}. The  spin projection probabilities $\Pimvar(\px,m)$ exhibit the  characteristic edge mode structure at the virtual edge $\mstar$ (\fig{fig:optimum:pops}). They also agree very well with those measured of the entanglement Hamiltonian (\fig{fig:QH:projection_probability_A}), with a fidelity $\mathcal{F}=98(1)\%$ for the momentum interval used for the optimization (\fig{fig:optimum:infidelity} and Methods).

We show in \fig{fig:optimum:energy} the measured ground and first-excited band spectra $E_n^{\text{var}}(\px)$ ($n=0,1$), along with the targeted pseudo-spectrum $\EA(\px)$. The modes with a pseudo-energy $\EA$ close to zero, which are significantly delocalized across $\mstar$, carry the most important information on spatial entanglement. They correspond to momenta $\px$ near $\pxstar$, for which the ground band $\Ezerovar(\px)$ reproduces well $\EA(\px)$.
Away from this zone, $\Ezerovar$   deviates from $\EA$, which can be attributed to (i)  for $\px\gtrsim\pxstar$, the discrete nature of the synthetic dimension, which leads to a quadratic dispersion once the system is fully polarized at  $m=0$ (ii) for $\px\lesssim-2J\hbar k$, the finite size of the synthetic dimension, which also leads to quadratic dispersion when polarized in  the lowest state $m=-J$. Such deviations would not occur for a continuous Hall system restricted to a semi-infinite half plane, for which the ground band $\EzeroBW(\px)$ is strictly linear. However, these deviations are also present for the BW ansatz of our synthetic system, as discussed in Methods.

\begin{figure}
\includegraphics[scale=0.89,trim=2.5mm 2mm 0 0]{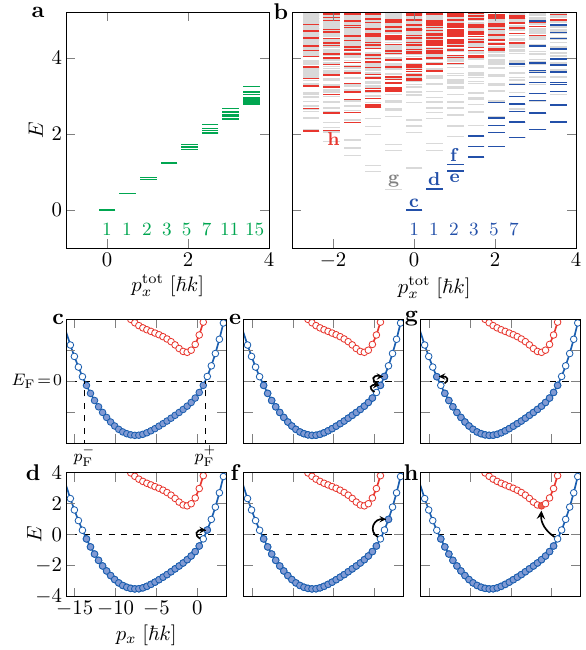}
    \phantomsubfloat{\label{fig:entanglement:entanglement_spectrum}}
    \phantomsubfloat{\label{fig:entanglement:many-body_spectrum}}
    \phantomsubfloat{\label{fig:entanglement:single-particle_c}}
    \phantomsubfloat{\label{fig:entanglement:single-particle_d}}
    \phantomsubfloat{\label{fig:entanglement:single-particle_e}}
    \phantomsubfloat{\label{fig:entanglement:single-particle_f}}
    \phantomsubfloat{\label{fig:entanglement:single-particle_g}}
    \phantomsubfloat{\label{fig:entanglement:single-particle_h}}
    \vspace{-2\baselineskip}
\caption{
\textbf{Many-body entanglement spectrum.}
\captionlabel{fig:entanglement:entanglement_spectrum} Many-body entanglement spectrum of a quantum Hall insulator computed from the measured single-particle pseudo-spectrum, for an ensemble of $\Nat=60$ atoms and for a system length $L_x=12.3\,k^{-1}$.
 \captionlabel{fig:entanglement:many-body_spectrum} Many-body spectrum of a system of $N_{\text{at}}^A=30$ fermions evolving in the optimal variational Hamiltonian, for the same system length.  We distinguish the full spectrum (levels in gray) from states only involving  excitations close to the Fermi point $\pFp$, either within the ground band (in blue) or among excited bands as well (in red). We indicate, for each total momentum $p_x^{\text{tot}}$, the number of states involving the ground band only. In b, the counting is limited to the range $0\leq\px<3\,\hbar k$ over which ground-band excitations are separated from higher band excitations.
 \captionlabel{fig:entanglement:single-particle_c}
 \captionlabel{fig:entanglement:single-particle_d}
 \captionlabel{fig:entanglement:single-particle_e}
 \captionlabel{fig:entanglement:single-particle_f}
 \captionlabel{fig:entanglement:single-particle_g}
 \captionlabel{fig:entanglement:single-particle_h}
 Occupation of momentum orbitals for a selection of states as indicated in b. The Fermi level $\EF=0$ is shown as a dashed line.
\label{fig:entanglement}
}
\end{figure}

\medskip
\noindent{\textbf{Many-body entanglement spectrum}}\\
\noindent Our work represents an important step towards the measurement of entanglement properties of quantum many-body systems. As a first validation of our approach, we show in Methods that the entanglement entropy is mapped to the thermodynamic entropy of a Fermi gas evolving under the optimal variational Hamiltonian.

We focus here on measuring the many-body entanglement spectrum of a fermionic Hall insulator. We consider an ensemble of non-interacting fermions filling the ground band of the optimal variational Hamiltonian. For a length $L_x=12.3\,k^{-1}$, the many-body ground state of $\Nat=60$ atoms is a Fermi sea with unit occupancy of momentum states $|\px|<\pF=15.2\,\hbar k$ of the ground band of the undeformed Hall system (\fig{fig:QH:dispersion_relation}). We show in \fig{fig:entanglement:entanglement_spectrum} the  entanglement spectrum describing the reduced system in subregion $A$, assuming a number of particles $N_{\text{at}}^A=30$. We obtain it by computing the energy of all possible occupancies of the orbitals $\psiA(\px)$ involved in the  entanglement Hamiltonian \eqP{eq:KA}, using the measured pseudo-spectrum $\EA(\px)$ (\fig{fig:QH:EA}). This entanglement spectrum exhibits  a chiral dispersion, with a number of states per value of the total momentum $p_x^{\text{tot}}$ characteristic of a chiral bosonic mode.

We compare it with the energy spectrum of $N_{\text{at}}^A$ fermions evolving in the optimal variational Hamiltonian $\KAopt$, with the same system length $L_x$. The ground state corresponds to a Fermi sea with unit filling over the interval $\pFm<\px<\pFp$, i.e. a Fermi energy $\EF=0$ (\fig{fig:entanglement:single-particle_c}). The many-body spectrum is obtained by computing the energy of all possible occupancies of the measured bandstructure $E_n^{\text{var}}(\px)$ (\fig{fig:optimum:energy}). The full spectrum, shown in gray in \fig{fig:entanglement:many-body_spectrum}, does not show chirality, due to the contribution of excitations around the two Fermi points $\pFm$ (\fig{fig:entanglement:single-particle_g}) and $\pFp$ (\fig{fig:entanglement}d,e,f). Excitations around the Fermi point $\pFp$ can be separated by exciting the system locally around $m=\mstar$, so that the overlap with orbitals around $\pFm$ practically vanishes. The corresponding excitation spectrum (blue levels in \fig{fig:entanglement:many-body_spectrum}) then exhibits the same structure than the entanglement spectrum, including the counting of states. Note that the spectrum also contains excitations to higher bands (red levels in \fig{fig:entanglement:many-body_spectrum} and \fig{fig:entanglement:single-particle_h}), which limits the range over which the entanglement spectrum can be identified.

The protocols outlined in this work can be readily transposed to a wide range of quantum simulators, such as lattice atomic gases, trapped ions, Rydberg atom arrays, and superconducting quantum circuits. To generate the entanglement Hamiltonian of a given many-body system, one simply needs to engineer a spatial deformation of the single-particle Hamiltonian -- following the procedure demonstrated here -- and add interactions. For integrable one-dimensional systems, the correspondence between the BW and entanglement Hamiltonians  can be established analytically \cite{peschel_density-matrix_1999,peschel_reduced_2009,dalmonte_entanglement_2022}. The relevance of the BW Hamiltonian has also been confirmed numerically for various lattice models \cite{dalmonte_quantum_2018,giudici_entanglement_2018}, including critical quantum spin chains \cite{mendes-santos_entanglement_2019} and a topological Haldane spin chain \cite{dalmonte_quantum_2018,zache_entanglement_2022}, and we are currently investigating theoretically its extension to fractional quantum Hall states. By directly engineering the entanglement Hamiltonian, we expect to gain experimental access to entanglement properties of many-body systems that would otherwise remain inaccessible, allowing for identification of topological order, critical behavior, or more generally the structure of correlations between particles.

We mention that, after completing the present work, we became aware of a related study in which the entanglement spectrum of a XXZ spin chain was measured in a trapped ion system  \cite{joshi_exploring_2023}.


\providecommand{\noopsort}[1]{}

\newcommand{\titlesupp}[1]{{\noindent\textbf{#1}}}

\bigskip
\titlesupp{Acknowledgements.} We thank Tanish Satoor for his contribution to the preparation of this project. We acknowledge insightful discussions with Marcello Dalmonte, Benoît Estienne, Michael Fleischhauer, Nicolas Regnault and Torsten Zache. We thank Jean Dalibard and Leonardo Mazza for discussions and careful reading of the manuscript. This work is supported by  European Union (grant TOPODY 756722 from the European Research Council) and Institut Universitaire de France. N.M. acknowledges support from DIM Quantip of région \^Ile de France.

\medskip
\titlesupp{Author contributions.}
All authors contributed to the set-up of the experiment, data acquisition, data analysis and the writing of the manuscript.

\medskip
\titlesupp{Competing interests.}
The authors declare no competing interests.

\medskip
\titlesupp{Data availability.} 
Source data, as well as other datasets generated and analysed during the current study, are available from the corresponding author upon request. 

\clearpage

\newcommand{\sectionmethods}[1]{
\medskip
\noindent\textbf{#1}
\smallskip\\
}

\noindent{\Large \textbf{Methods}}

\begin{figure}
\includegraphics[scale=1,trim=3mm 0 0 0]{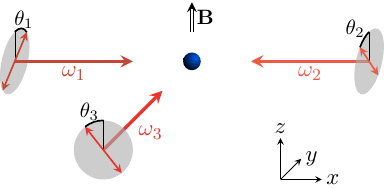}
\caption{
\textbf{Scheme of the laser configuration.} A pair of laser beams counter-propagating along $x$, labeled 1 and 2, induces two-photon Raman transitions between successive magnetic projection states $m$, for a frequency difference $\omega_2-\omega_1$ close to the Zeeman splitting $\omZ$ induced by an external magnetic field. The algebra of the spin coupling is determined by the orientations $\theta_1$ and $\theta_2$ of the linear polarizations of the two beams. A third laser beam propagating along $y$ generates a quadratic Zeeman light shift $Q_3 J_z^2$.  We use a polarization angle $\theta_3=0$ or $\pi/2$ depending on the desired sign of $Q_3$.
\label{fig:laser_scheme}
 }
\end{figure}

\sectionmethods{Light-induced spin-orbit coupling}
\noindent The spin dynamics is induced by two-photon optical transitions, which are driven by  a pair  of laser beams counter-propagating along the $x$-axis, with their linear polarizations parametrized by the angles $\theta_1$, $\theta_2$ with respect to the $z$-axis (\fig{fig:laser_scheme}). The laser  frequencies $\omega_1$ and $\omega_2$ are set close to the optical resonance of wavelength $\lambda=\SI{626.1}{\nano\meter}$. In addition, we apply a bias magnetic field $B=\SI{0.956(3)}{G}$ along $z$, which induces a Zeeman splitting of frequency $\omZ=\SI{1.662(5)}{\mega\hertz}$. Since this splitting is much larger than the characteristic energy scale of atomic motion (determined by the recoil energy $\Erec=h\times\SI{3.14}{\kilo\hertz}$), spin transitions only take place when the frequency difference $\omega_2-\omega_1$ between the two lasers is close the Zeeman splitting $\omZ$. A Raman transition $m\rightarrow m+1$ then occurs via the absorption (resp. emission) of one photon from laser 2 (resp. laser 1), so that the atom gets a velocity kick $-2\vrec$ along $x$. Within the rotating wave approximation (RWA),  the atom dynamics is governed by the Hamiltonian
\[
 H=\frac{(\px-2\hbar k J_z)^2}{2M}+V-\hbar\delta J_z,
\]
where we introduce the detuning $\delta=\omega_2-\omega_1-\omZ$, and
\begin{align*}
 \frac{V}{V_0}=\frac{1}{2(J+1)(2J+1)}\Big[&-(2J+3)\sin(\theta_2-\theta_1)J_x
 \\
 &+\sin(\theta_1+\theta_2)\{J_x,J_z\}\\
 &+\left(2-3\cos^2\theta_1-3\cos^2\theta_2\right)J_z^2\Big].
\end{align*}
Here, $V_0=3\pi c^2\Gamma I/(2\omega_0^3\Delta)$ is the light shift for Clebsch-Gordan coefficients equal to unity, $\Gamma=\SI{0.85}{\micro\second^{-1}}$ is the transition linewidth, $\omega_0$ is its resonant frequency, $\Delta=-2\pi\times\SI{10.8(1)}{\giga\hertz}$ is the laser detuning from resonance.

To generate the undeformed quantum Hall system given in \eqP{eq:H_QH} of the main text, we choose $\theta_1=-\theta_2=\acos(1/\sqrt3)=\SI{54.7}{\degree}$, so that the atom-light coupling reduces to a linear spin operator
\[
 \frac{V}{V_0}=\frac{\sqrt 8(2J+3)}{6(J+1)(2J+1)}J_x.
\]

The spin coupling $\{m^\star-J_z,J_x\}$ involved in the variational Hamiltonian of \eqP{eq:HAvar} in the main text is obtained for a different choice of polarizations. Keeping the polarization $\theta_1=\acos(1/\sqrt3)$, the polarization $\theta_2$ should satisfy $(2J+3)\sin(\theta_2-\theta_1)=\sin(\theta_1+\theta_2)$, leading to
\[
 \theta_2=\atan\left(\sqrt 2\frac{J+2}{J+1}\right)=\SI{57.5}{\degree}.
\]
The atom-light coupling then reads
\begin{align*}
 V&= -\frac{\hbar\Omegavar}{2J}(J_+J_z+J_z J_-)+Q_{12} J_z^2, \\
 \hbar\Omegavar &=-\sqrt{\frac{2}{3(3J^2+10J+9)}}\frac{J(2J+3)}{(J+1)(2J+1)}V_0\\&\simeq -0.048\,V_0,\\
Q_{12}&=\frac{2J+3}{(3J^2+10J+9)(J+1)(2J+1)}V_0\\&\simeq4.4\times10^{-4}\,V_0.
\end{align*}

We also use a third laser beam that propagates along the $y$-axis and is significantly off-resonant with respect to the other two beams. We use its tensor light shift to  generate a quadratic Zeeman shift $Q_3 J_z^2$, which can be tuned independently from the other spin couplings. The total quadratic Zeeman shift amplitude then writes $\Qvar=Q_{12}+Q_3$.

\sectionmethods{Preparation of ground-band momentum states}
\noindent Our experiments consist of the preparation and characterization of thermal ensembles of $^{162}$Dy atoms subjected to a light-induced spin-orbit coupling and evolving in the  ground energy band. We use standard laser cooling techniques to prepare an atomic gas of $N_{\text{at}}=2.4(3)\times10^4$ atoms in a crossed optical dipole trap with a temperature of $T=\SI{350(50)}{\nano\kelvin}$, which corresponds to an rms momentum width of $\sigma_{\px}=1.5(1)\,\hbar k$.

At the end of the cooling sequence, the atoms are polarized in the ground state $m=-J$. We then turn off the dipole trap, and ramp up the Raman laser intensities over a duration of $\SI{50}{\micro\second}$. At this stage, we use a large negative detuning $\delta\simeq-80\,\Erec/\hbar$, so that spin transitions are off resonant.
We interpret our experiments in a moving reference frame, which is defined by the velocity $\vx^*=-\delta/(2k)$ with respect to the laboratory frame, so that the  detuning $\delta$ is exactly compensated by the Doppler effect. In this frame, the initial state  corresponds to an initial momentum $\px\simeq-40\,\hbar k$.

\begin{figure}
\includegraphics[scale=0.87,trim=3mm 0 0 0]{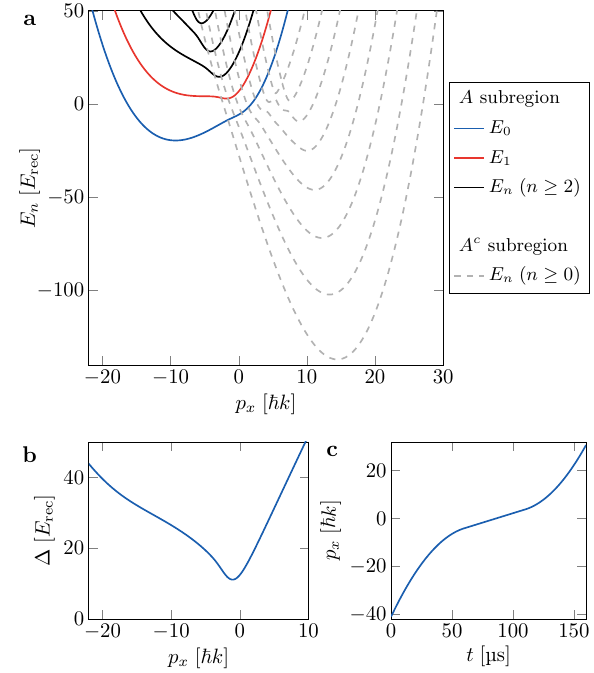}
    \phantomsubfloat{\label{fig:ramp:spectrum}}
    \phantomsubfloat{\label{fig:ramp:gap}}
    \phantomsubfloat{\label{fig:ramp:detuning}}
    \vspace{-2\baselineskip}
\caption{
\textbf{Adiabatic state preparation.} 
\captionlabel{fig:ramp:spectrum} Theoretical dispersion relation of the variational Hamiltonian for $\Omegavar=15\,\Erec/\hbar$, $\bvar=-10\,\Erec$ and $\Qvar=-0.5\,\Erec$. The solid (resp. dashed) lines correspond to the bandstructure for the $A$ (resp. $A^c$) subregion. In the absence of imperfection, these two sectors are decoupled.
\captionlabel{fig:ramp:gap} Energy gap $\Delta=E_1-E_0$ as a function of momentum.
\captionlabel{fig:ramp:detuning} Time variation of the momentum $\px$, controlled via the detuning between the Raman lasers, used to ramp it adiabatically across the ground band for $\Omegavar=13.9(8)\Erec/\hbar$. The ramp speed is reduced around $\px=0$, where the gap to the first energy gap is minimal. In order to prepare a given momentum state $\px$, we interrupt the ramp at the corresponding time.
\label{fig:ramp}
 }
\end{figure}

We then adiabatically ramp up the momentum $\px$ by varying the  detuning $\delta$, which induces an inertial force in the moving frame.
To ensure adiabatic dynamics, we choose a slow detuning ramp profile. For the undeformed quantum Hall system, for which the  energy gap between the ground and first excited bands is quasi-uniform in the bulk, we simply use a linear ramp.

We show in \fig{fig:ramp:spectrum} the theoretical bandspectrum for the deformed variational Hamiltonian, which consists of two decoupled bandstructures within each subregion $A$ and $A^c$. The gap between the ground and first excited bands of $A$ exhibits a significant variation with $\px$, taking a minimum value close to $\px=0$ (\fig{fig:ramp:gap}). Therefore, we use a variable ramp speed, shown in \fig{fig:ramp:detuning}, to ensure adiabaticity. Note that the ground band crosses several bands of the $A^c$ bandstructure. While we aim to cancel the coupling between $A$ and $A^c$, a residual coupling remains, so that the ramp must be fast enough to ensure diabatic dynamics at these band crossings. We have optimized the ramp shape for each value of $\Omegavar$ to ensure adiabatic dynamics within $A$, while minimizing the transfer of atoms to $A^c$, as described in the next section.  Note that collisions between atoms, which can induce a redistribution of momentum and atom loss due to dipolar relaxation, occur on a 10-ms time scale, much longer than the typical ramp duration $\sim\SI{100}{\micro \second}$.

\begin{figure}
\includegraphics[scale=0.89,trim=2mm 0 0 0]{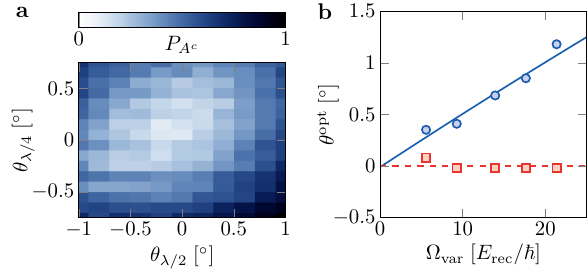}
    \phantomsubfloat{\label{fig:decoupling_waveplates}}
    \phantomsubfloat{\label{fig:decoupling:beyondRWA}}
    \vspace{-2\baselineskip}
 \caption{\textbf{Decoupling between $A$ and $A^c$.}
 \captionlabel{fig:decoupling_waveplates} Fraction of atoms in $A^c$ as a function of the orientation of half- and quarter-waveplates, after an adiabatic ramp across the ground energy band.
 \captionlabel{fig:decoupling:beyondRWA} Variation of the optimal waveplate orientations measured as a function of the Raman coupling $\Omegavar$ (blue disks: half-waveplate, red squares: quarter-waveplate). The lines are the theoretical value expected from the deviation from RWA given in  \eqP{eq:deltatheta2} (blue solid line: half-waveplate, red dashed line: quarter-waveplate).
 }
\end{figure}

\sectionmethods{Decoupling between $A$ and $A^c$}
The spin coupling $\{m^\star-J_z,J_x\}=(J_+J_z+J_zJ_-)/2$ present in the variational Hamiltonian $\KAvar$ does not couple the subregions $A$  and $A^c$. However, in our experiments, we observe that when we increase the momentum quasi-adiabatically across the ground band of the deformed variational Hamiltonian, the atoms exhibit a residual probability $P_{A^c}$ of occupying $A^c$, while they should ideally end up fully polarized in $m=0$. We attribute this behavior to a defect in the polarization of the Raman lasers. The decoupling between $A$ and $A^c$ relies on a fine tuning of polarization, which we achieve using a pair of motorized half- and quarter-waveplates. By tuning the waveplate orientation with $\SI{0.2}{\degree}$ sensitivity, we are able to achieve a minimum value of $P_{A^c}\simeq8\%$ (\fig{fig:decoupling_waveplates}). To reduce the residual population in $A^c$ further, we could shorten the ramp duration of the laser detunings used to prepare a given momentum state. However, this would result in a degradation of the adiabaticity of the state preparation. Therefore, the chosen ramp duration and value of $P_{A^c}$ result from a compromise.

We have observed that the half-waveplate orientation needs to be adjusted as a function of the Raman coupling strength $\Omegavar$ (\fig{fig:decoupling:beyondRWA}). We attribute this effect to a deviation from the RWA. To lowest order in the inverse Zeeman splitting $1/\omZ$, we compute the residual coupling between the states $m=0$ and $m=1$ due to the first correction to the RWA \cite{goldman_periodically_2015}. This coupling can be compensated by rotating the linear polarization of laser 2 by an angle
\begin{equation}\label{eq:deltatheta2}
 \delta\theta_{2}=\frac{2J+2}{2J+3}\frac{\Omegavar}{\omZ}.
\end{equation}
This can be achieved by rotating the half-waveplate, while the  quarter-waveplate does not need to be rotated.  The measured variations of the optimal waveplate positions, which are shown in \fig{fig:decoupling:beyondRWA}, are in good agreement with this prediction.

We have also observed a similar effect when the laser 3 is used to control the quadratic Zeeman field $\Qvar$. We compensate for the induced coupling between $A$ and $A^c$ by further tuning  the half-waveplate  for each value of $\Qvar$.

\begin{figure}
\includegraphics[scale=0.9,trim=3mm 8mm 0 0]{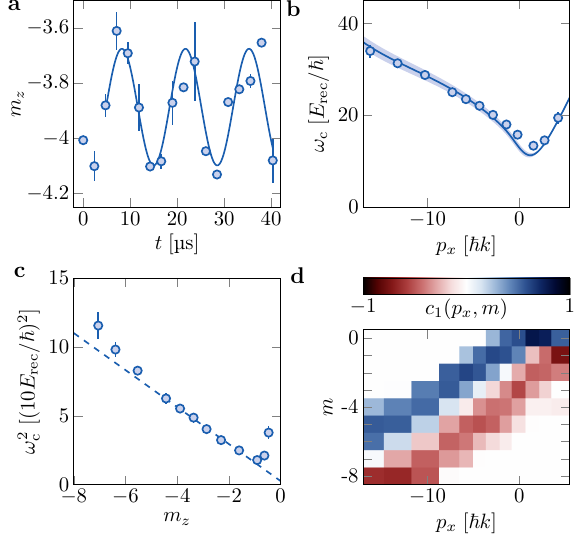}
    \phantomsubfloat{\label{fig:cyclotron:evolution}}
    \phantomsubfloat{\label{fig:cyclotron:omegac}}
    \phantomsubfloat{\label{fig:cyclotron:omegac2}}
    \phantomsubfloat{\label{fig:cyclotron:c1}}
 \caption{\textbf{Cyclotron frequency measurements.}\label{fig:cyclotron}
\captionlabel{fig:cyclotron:evolution} Example of time evolution of the mean spin projection $m_z$ subsequent to a velocity kick, for a momentum $\px=-5.8(2)\,\hbar k$ and  variational parameters $\Omegavar=13.9(8)\Erec/\hbar$, $\bvar=0.0(3)$ and $\Qvar=-0.1(2)\,\Erec$. The solid line is a sine fit that yields the cyclotron frequency $\omc$. We do not take into account hold times $t\leq\SI{2.5}{\micro\second}$ during which the Raman
detuning relaxes after quenching.
 \captionlabel{fig:cyclotron:omegac} Cyclotron frequency $\omc$ as a function of momentum $\px$ (blue circles). The blue line is the expected gap variation.
\captionlabel{fig:cyclotron:omegac2} Same data plotted as $\omc^2$ against the mean spin projection $m_z$. The dashed line is a linear guide to the eye. The approximate linear variation in the range $-6<m_z<-2$ is consistent with an inverse mass $1/M_m$ varying linearly with $m$.
\captionlabel{fig:cyclotron:c1} Wavefunction amplitude $c_1(\px,m)$ in the first excited band, deduced from the dynamics of individual $m$ states.
}
\end{figure}

\sectionmethods{Cyclotron frequency measurements}
To calibrate the Raman couplings $\Omega$ and $\Omegavar$, we measure the energy gap between the ground and first excited bands, using a common protocol.  We present here the case of the deformed quantum Hall system $\HAvar$ (\fig{fig:cyclotron}).

To induce a coherent mixing with excited bands, we quench the Raman detuning after preparing a given momentum state in the ground band. This imparts a velocity kick to the atoms with a magnitude of approximately $0.5\,\vrec$, which is small enough to restrict the excitation to the first excited band. This results in a quasi-harmonic dynamics with a frequency of $\omc=(E_1-E_0)/\hbar$, as shown (\fig{fig:cyclotron:evolution}).
In the case of the deformed Hall system, we observe a strong variation of $\omc$ with $\px$  (\fig{fig:cyclotron:omegac}). This can be attributed to the strong dispersion of matrix elements of the spin coupling $\{m^\star-J_z,J_x\}$, which can be interpreted  as a position-dependent mass $M_m\propto 1/(m^\star-m)$ for the dynamics along $m$. As in the case of a Hall system with anisotropic mass \cite{ciftja_cyclotron_2017}, we expect the cyclotron frequency to scale as the inverse of the geometric mean $\sqrt{M M_m}$ of the masses along $x$ and $m$, in agreement with our measurements (\fig{fig:cyclotron:omegac2}).

In order to calculate the entropy density $s_{\text{var}}(m)$ (see last section of Methods), we rely on the spin projection probabilities $\Pi_1(\px,m)$ of the momentum state $\px$ in the first excited band, as well as other previously measured quantities. To obtain $\Pi_1(\px,m)$, we analyze the cyclotron dynamics of the spin projection probabilities, which are expected to evolve according to
\[
 \Pi(m,t)=|c_0(\px,m)+\epsilon \,c_1(\px,m)\E^{-\I\omc t}|^2,
\]
where $\epsilon$ is a weak excitation amplitude  and we neglect terms in $\epsilon^2$.
Here, we define the spin projection amplitudes $c_n(\px,m)$ for the bands $n=0,1$, such that the projection probabilities are given by $\Pi_n(\px,m)=|c_n(\px,m)|^2$. Since the Hamiltonian is real, we can assume that the oscillation amplitudes are real for all bands, and that $c_0(m)\geq0$ for the ground band. We rewrite the spin projection probabilities as  (omitting the label $\px$ for simplicity)
\[
 \Pi(m,t)=\Pi_0(m)+A(m)\cos(\omc t+\phi),
\]
with the oscillation amplitude $A(m)=2|\epsilon| c_0(m)c_1(m)$. The excited-band amplitude then writes
\[
 c_1(m)=\frac{A(m)}{2|\epsilon|\sqrt{\Pi_0(m)}}.
\]
It can thus be obtained
from the measurement of the oscillation amplitude $A(m)$, combined with the ground-state projection probability $\Pi_0(m)$. We show in \fig{fig:cyclotron:c1} the excited-band amplitude $c_1(\px,m)$ as a function of $\px$ deduced from this protocol. Notably, we find that the amplitude $c_1$ takes both positive and negative values. This behavior is expected for the first excited band, according to a generalization of the node oscillation theorem  for spin systems \cite{ribeiro_exact_2008}. By taking the square of the amplitude $c_1(\px,m)$, we obtain the spin projection probability $\Pi_1(\px,m)$.

\sectionmethods{BW theorem for a continuous Hall system}
We demonstrate that the BW theorem holds true for an ideal quantum Hall system that is described by Landau levels. We consider an electron that evolves in a 2D plane $xy$ and is subjected to a perpendicular magnetic field $B$. By adopting the Landau gauge, its dynamics is described by the Hamiltonian
\[
H_{xy}=\frac{(p_x-eBy)^2}{2M}+\frac{p_y^2}{2M}.
\]
This Hamiltonian is invariant upon translations along $x$, so that the momentum $p_x$ is conserved. For every $p_x$, the $y$ dynamics corresponds to a harmonic oscillator of cyclotron frequency $\omc=eB/M$, centered on $y_{p_x}=p_x/eB$. We introduce the ladder operator $a_{p_x}=[(y-y_{p_x})/\ell+\I p_y\ell/\hbar]/\sqrt 2$, with $[a_{p_x},a_{p_x}^\dagger]=1$. Additionally, we define the cyclotron length $\ell=\sqrt{\hbar/eB}$. The Hamiltonian can be rewritten as $H=\hbar\omc\sum_{p_x}(a_{p_x}^\dagger a_{p_x}+\tfrac{1}{2})$, and its spectrum consists of flat bands $E_n(p_x)=\hbar\omc(n+\tfrac{1}{2})$, the so-called Landau levels. The eigenstates of the ground band are characterized by the wavefunctions
\begin{equation}\label{eq:psip_LLL}
 \psi_{p_x}(x,y)=\frac{\E^{\I p_x x/\hbar}}{\sqrt{L_x}}\frac{\E^{-(y-y_{p_x})^2/2\ell^2}}{\pi^{1/4}\sqrt{\ell}}.
\end{equation}

We now consider the spatial partition of the system between a subregion $A$ defined by $y<0$ and its complement. The probability $P_A$ of projection in $A$ writes $P_A(p_x)=\erfc(p_x\ell/\hbar)/2$, where $\erfc$ is the complementary error function. This formula allows us to derive an analytical expression for the pseudo-spectrum as follows:
\begin{equation}\label{eq:EA_continuous}
 \EA(p_x)=\frac{\erfc(-p_x\ell/\hbar)}{\erfc(p_x\ell/\hbar)}.
\end{equation}
The entanglement Hamiltonian is then given by
\[
 \KA=\sum_{p_x}\EA(p_x)n(\px),
\]
where $n(\px)$ is the occupation number in the state $\ket{\psi_{p_x}^A}$, obtained by restricting the state $\ket{\psi_{p_x}}$ to the subregion $A$ and renormalizing as $\psi_{p_x}^A(x,y)=\psi_{p_x}(x,y)/\sqrt{P_A}$, where we only consider $y<0$.

We compare this expression of $\KA$ with the BW ansatz, which can be formulated as
\begin{equation}\label{eq:KABW_LLL}
\KABW = \frac{4}{\sqrt{\pi}}\left\{\frac{-y}{\ell}, \frac{H_{xy}-\Ezero}{\hbar\omc}\right\}.
\end{equation}
While the energy offset $\Ezero$  is irrelevant for the particle dynamics governed by $H_{xy}$,  it has a non-trivial role in the dynamics of $\KABW$. The choice of its value is crucial, and we will discuss this in detail later in this section.

In order to demonstrate that this Hamiltonian leaves the subregion $A$ invariant, we analyze the probability current along $y$, which is given by
\begin{align*}
 j_y&=\frac{1}{\hbar}\Imag \psi^*[y,\KABW]\psi,\\
 &=\frac{-2}{\sqrt{\pi}\ell\omc M}y\Imag \psi^*\partial_y\psi
\end{align*}
for a wavefunction $\psi$. Assuming a smooth wavefunction, the current  vanishes for $y=0$, which shows that particles cannot cross the line $y=0$.

We can rewrite the Hamiltonian $\KABW$ in terms of ladder operators, which gives us a new expression
\[
 \KABW = \frac{-4}{\sqrt{\pi}}\sum_{p_x}\left\{\frac{a_{p_x}+a_{p_x}^\dagger}{\sqrt2}+\frac{y_{p_x}}{\ell}, a_{p_x}^\dagger a_{p_x}+\frac{1}{2}-\frac{\Ezero}{\hbar\omc}\right\}.
\]
Rewritting it in normal order, we get
\begin{align}
 \KABW = \frac{-4\sqrt{2}}{\sqrt{\pi}}\sum_{p_x}&a_{p_x}^{\dagger2} a_{p_x}+a_{p_x}^\dagger a_{p_x}^2\nonumber\\
 &+\frac{p_x\ell }{\sqrt 2 \hbar}\left(2a_{p_x}^\dagger a_{p_x}+1-2\frac{\Ezero}{\hbar\omc}\right)\nonumber\\
 &+\frac{\hbar\omc-\Ezero}{\hbar\omc}(a_{p_x}^\dagger +a_{p_x}).\label{eq:KABW_normal_order}
\end{align}
The ground-band state $\ket{\psi_{p_x}^A}$  is  characterized by $a_{p_x}\ket{\psi_{p_x}^A}=0$, and it is also an eigenstate of $\KABW $ provided that the last term of \eqP{eq:KABW_normal_order} vanishes, i.e. for an energy offset $\Ezero=\hbar\omc$. The corresponding eigenenergy then reads
\begin{equation}\label{eq:E0_BW}
 \EzeroBW(p_x)=\frac{4\ell}{\sqrt\pi \hbar}p_x,
\end{equation}
which corresponds to the linear expansion of the entanglement spectrum $\EA(p_x)$ given in \eqP{eq:KABW_LLL} in the main text, evaluated around $p_x=0$.

\begin{figure}
\includegraphics[scale=1,trim=5mm 0 0 0]{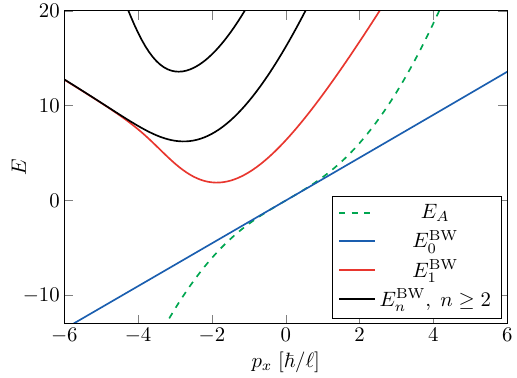}
\caption{
Energy spectrum of the Bisognano-Wichmann Hamiltonian $\KABW$ associated with a continuous quantum Hall system (solid lines). The dashed green line is the actual entanglement spectrum given by \eqP{eq:EA_continuous}.
\label{fig:spectrum_BW_QH}
}
\end{figure}

In addition to its ground band, the Hamiltonian $\KABW$ also features higher energy bands $E_n^{\text{BW}}$, with $n\geq1$.  The eigenstates of excited bands are not simply related to those of the Hamiltonian $H_{xy}$. The complete bandstructure calculated numerically is shown in  \fig{fig:spectrum_BW_QH}.

\begin{figure}
\includegraphics[scale=0.9,trim=3mm 8mm 0 0]{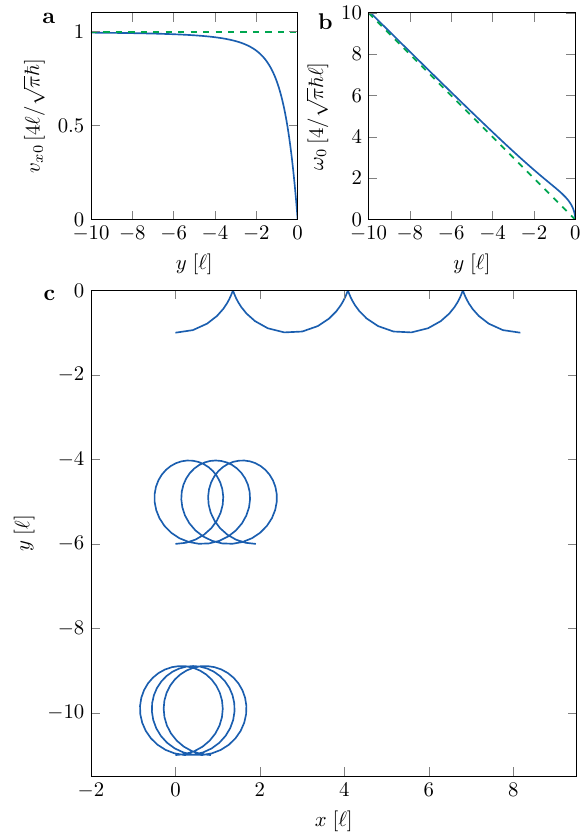}
    \phantomsubfloat{\label{fig:classical_trajectories:velocity}}
    \phantomsubfloat{\label{fig:classical_trajectories:frequency}}
    \phantomsubfloat{\label{fig:classical_trajectories:orbits}}
 \caption{\textbf{Classical trajectories of the BW Hamiltonian.}\label{fig:classical_trajectories}
\captionlabel{fig:classical_trajectories:velocity} Velocity $v_{x0}$ as a function of $y$ so that the trajectory reduces to a drift along $x$. The dashed green line is the group velocity of the ground band of $\KABW$.
 \captionlabel{fig:classical_trajectories:frequency} Oscillation frequency for small deviations around the solution of constant velocity. The dashed green line is the expected frequency assuming it is proportional to the local deformation factor of $\KABW$. 
\captionlabel{fig:classical_trajectories:orbits} Examples of orbits governed by $\KABW$, for different initial positions $y(t=0)$ and velocities $v_x(t=0)$, and $x(t=0)$, $v_y(t=0)=0$. The evolution time corresponds to three periods of the oscillatory part of the dynamics.
}
\end{figure}

To provide a comprehensive picture of the dynamics induced by $\KABW$, we also consider classical particle trajectories. However, since the Hamiltonian $\KABW$ is cubic in $\px$, $p_y$ and $y$, the Newton equations are non linear and must be solved numerically. We always find that the velocity dynamics are periodic, meaning that the trajectories can be decomposed into a periodic orbit and a drift of constant velocity.  For each $y$ position, there exists a single velocity  $v_{x0}(y)$ for which the dynamics reduces to a constant drift along the $x$ direction (\fig{fig:classical_trajectories:velocity}). This velocity tends to match the slope $4/\sqrt\pi\ell/\hbar$ of the ground band of $\KABW$ for $|y|\gg1$. We also find that perturbations around this solution correspond to an harmonic oscillation of frequency $\omega_{0}(y)$ (\fig{fig:classical_trajectories:frequency}). 
For $|y|\gg1$, this frequency is proportional to the local deformation factor $|y|$ involved in the definition of $\KABW$.  

To visually explore the classical trajectories, we show in \fig{fig:classical_trajectories:orbits} three orbits computed for different choices of the initial $y$ position. Each orbit is represented over three oscillations of the periodic part of the dynamics.  They correspond well to the schematics of \fig{fig:scheme:KA} in the main text. Specifically, we observe skipping orbits close to $y=0$ and quasi-closed cyclotron orbits for $|y|\gg\ell$.

\sectionmethods{BW ansatz of our synthetic Hall system}
We can apply the Bisognano-Wichmann ansatz to our synthetic Hall system, for which the partition cut is defined as a line $m=\mstar$, as
\begin{equation}\label{eq:KABW_synthetic}
\KABW = \frac{4}{\sqrt{\pi}}\left\{\frac{m^\star-J_z}{\ell_m}, \frac{H-\Ezero}{\hbar\omc}\right\},
\end{equation}
where we recall the expression of the Hamiltonian
\[
 H=\frac{(\px-2\hbar k J_z)^2}{2M}-\hbar\Omega J_x+Q J_z^2.
\]
The magnetic length $l_m$, the cyclotron frequency $\omc$ and the energy offset $\Ezero$ can be determined by mapping the synthetic system onto a standard Hall system described by the Landau Hamiltonian $H_{xy}$.

\begin{figure}
\includegraphics[scale=1,trim=5mm 0 0 0]{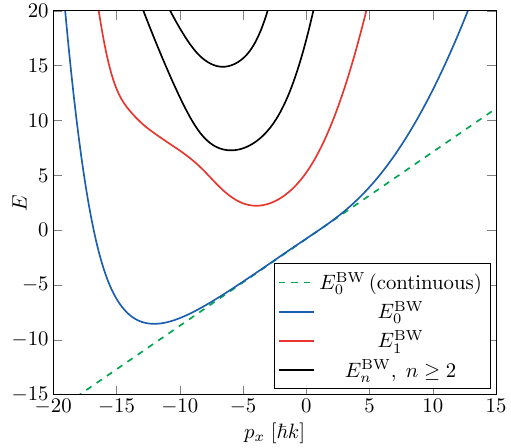}
\caption{
Energy spectrum of the Bisognano-Wichmann Hamiltonian $\KABW$ associated with our synthetic quantum Hall system (solid lines). The dashed green line is the linear dispersion expected for the equivalent continuous system.
\label{fig:spectrum_BW_synthetic}
}
\end{figure}

In the middle of the bulk mode region, where $\px\simeq0$, the spin is almost polarized along $x$. As a result, we can approximate  the commutator $[J_z,J_y]=-\I J_x$ by a constant, as $J_x\simeq J$.  We can define the operator $P_m=-\hbar J_y/J$ to be the momentum conjugate to the spin projection $m$, as it satisfies the relationship $[J_z,P_m]\simeq \I\hbar$. By expanding $J_x=\sqrt{J(J+1)-J_y^2-J_z^2}$ to second order in $J_y$, $J_z$, we get the quadratic Hamiltonian
\[
 H\simeq\frac{(\px-2\hbar k J_z)^2}{2M}+\frac{J\Omega}{2\hbar}P_m^2-\hbar\Omega\left(J+\frac{1}{2}\right),
\]
where the quadratic term in $J_z^2$ from the expansion of $J_x$ is compensated by  $Q J_z^2$. We recognize, up to a constant, the Landau Hamiltonian in which the spin projection $m$ plays the role of the dimension $y$, with a cyclotron frequency and a magnetic length
\[
\omc=\sqrt{8J \Omega\Erec/\hbar},\quad \ell_m=\left(\frac{J \hbar\Omega}{8\Erec}\right)^{1/4}.
\]
For the coupling $\Omega=3.6(1)\,\Erec/\hbar$ used in our experiments, these expressions give $\omc=15.2(2)\,\Erec/\hbar$ and $\ell_m=1.38(1)$, which is close to the measured values  $\omc=14.2(6)\,\Erec/\hbar$  and $\ell_m=1.53(5)$. Finally, the energy offset $\Ezero$ involved in the BW ansatz of \eqP{eq:KABW_synthetic} writes
\[
 \Ezero=-\hbar\Omega\left(J+\frac{1}{2}\right)-\hbar\omc.
\]
The bandstructure of the BW Hamiltonian of our synthetic Hall system is shown in  \fig{fig:spectrum_BW_synthetic}. Its ground band displays a wide momentum range with a linear dispersion that matches the strictly linear dispersion of the equivalent continuous Hall system. The deviation occurring for $\px\gtrsim 0$ can be attributed to the discrete nature of the synthetic dimension, since the dispersion becomes ballistic when the system is polarized in $m=0$. Similarly, the deviation for $\px\lesssim-10\,\hbar k$ is due to the polarization of the system in $m=-J$. A similar behavior is observed for the variational Hamiltonian studied experimentally (\fig{fig:optimum:energy} in the main text).

\sectionmethods{Equivalence Zeeman field/change of frame}
In our experiments, we implement a set of Hamiltonians
\[
\HAvar
=\frac{(\px-2\hbar k J_z)^2}{2M}-\frac{\hbar\Omegavar}{2J}(J_+J_z+\text{h.c.})+\Qvar J_z^2,
\]
which belong to the family of variational Hamiltonians given in \eqP{eq:HAvar} in the main text, with a Zeeman field $\bvar=0$.

We use the relation
\[
 \frac{(\px-2\hbar k J_z)^2}{2M}+\bvar J_z=\frac{(\px-\delta p-2\hbar k J_z)^2}{2M}+\frac{\delta p}{M}\px-\frac{\delta p^2}{2M},
\]
with $\delta p=\bvar/(2\vrec)$, to show that the application of a Zeeman field is equivalent to considering the system in a different reference frame and with a different value of momentum. Specifically, the properties of the variational Hamiltonian with $\bvar\neq0$ can be related to those for $\bvar=0$, according to
 \begin{align*}
 E_n^{\text{var}}(\px)|_{\bvar}&=E_n^{\text{var}}(\px-\delta p)|_{\bvar=0}+\frac{\delta p}{M}\px-\frac{\delta p^2}{2M},\\
 \ket{\psi_A^{\text{var}}(\px)}\big|_{\bvar}&=\ket{\psi_A^{\text{var}}(\px-\delta p)}\big|_{\bvar=0}.
 \end{align*}
Thus, it is  sufficient to realize a restricted class of variational Hamiltonians with $\bvar=0$, and derive the generic case $\bvar\neq0$ from it.

\sectionmethods{Relative entropy}
To optimize the variational Hamiltonian, we seek to minimize the relative entropy $S(\rho_A||\rhovar)=\Tr[\rho_A(\log\rho_A-\log\rhovar)]$ between the reduced density matrix $\rho_A$ and a thermal state $\rhovar$ of $\HAvar$, with temperature $\Tvar$ and chemical potential $\muvar$.
Defining the energy bands $E_n^{\text{var}}(\px)$ of the dimensionless Hamiltonian $\KAvar=(\HAvar-\muvar)/\kB\Tvar$, the relative entropy can be expressed as
\begin{align*}
 S(\rho_A||\rhovar) &=-S(\rho_A)\\&+\sum_{n,\px}\frac{\mathcal{F}_n(\px)E_n^{\text{var}}(\px)}{1+\E^{-\EA(\px)}}+\log(1+\E^{-E_n^{\text{var}}(\px)}),
\end{align*}
where $\mathcal{F}_n(\px)$ is the squared overlap between the momentum state $\px$ of the $n^{\text{th}}$ band and the state $\ket{\psi_A(\px)}$.

We have checked numerically that neglecting the bands $n\geq2$ does not affect the values of relative entropy significantly. Thus, we limit the summation to the ground and first excited bands $n=0,1$, and express the overlap of the first excited band in terms of the ground-band overlap $\mathcal{F}_0$ (denoted $\mathcal{F}$ hereafter), as $\mathcal{F}_1=1-\mathcal{F}$. This leads to
 \begin{align*}
S(\rho_A||\rhovar)=&\sum_{\px}\frac{E_0^{\text{var}}-E_A}{1+\E^{\EA}}+\log\frac{1+\E^{-E_0^{\text{var}}}}{1+\E^{-\EA}}\\
&+(1-\mathcal{F})\frac{E_1^{\text{var}}-E_0^{\text{var}}}{1+\E^{\EA}}+\log(1+\E^{-E_1^{\text{var}}})
\end{align*}
where all quantities depend on $\px$.
The summation in the previous equation can be divided in three blocks – the first two terms and the third and fourth terms respectively. Each of these blocks is positive, and they all cancel out when $E_0^{\text{var}}=E_A$, $\mathcal{F}=1$, and $E_1^{\text{var}}\rightarrow\infty$.

In our experiments, we obtain the value of the relative entropy from (i) the ground band dispersion $\Ezerovar(\px)$, which is obtained by integrating  the mean velocity $\langle v_x\rangle$, (ii) the first excited band dispersion $\Eonevar(\px)$, which we derive through the gap $\Delta(\px)$ measured by exciting cyclotron dynamics, and (iii) the fidelity $\mathcal{F}(\px)$. For the latter, we use the fact that, since  the variational Hamiltonian is real, the states of the ground band $|\psi_A^{\text{var}}(\px)\rangle$ can be expanded over spin projection states with real and positive coefficients, as
\[
 \langle x,m|\psi_A^{\text{var}}(\px)\rangle=\E^{\I\px x/\hbar}\sqrt{\Pivar(\px,m)}.
\]
A similar argument can be applied to the undeformed quantum Hall system, so that the fidelity can be expressed as
\begin{align*}
 \mathcal{F}(\px)&=|\langle\psi_A(\px)|\psi_A^{\text{var}}(\px)\rangle|^2\\
 &=\left(\sum_{m=-J}^0\sqrt{\Pi_A(\px,m)\Pivar(\px,m)}\right)^2,
\end{align*}
which enables us to determine it from measurements of spin projection probabilities.

In practice, we compute the relative entropy over a finite momentum interval $-10\hbar k<\px<5\hbar k$. The choice of the upper bound is based on the maximum momentum for which the states $\ket{\psi_A(\px)}$ can be accurately determined (\fig{fig:QH:projection_probability_A} in the main text). We choose the lower limit of the interval to maximize the momentum range used for the minimization process, while ensuring that the fidelity remains close to 1 over the entire interval (\fig{fig:optimum:infidelity} in the main text).

\begin{figure}
\includegraphics[scale=0.9,trim=3mm 8mm 0 0]{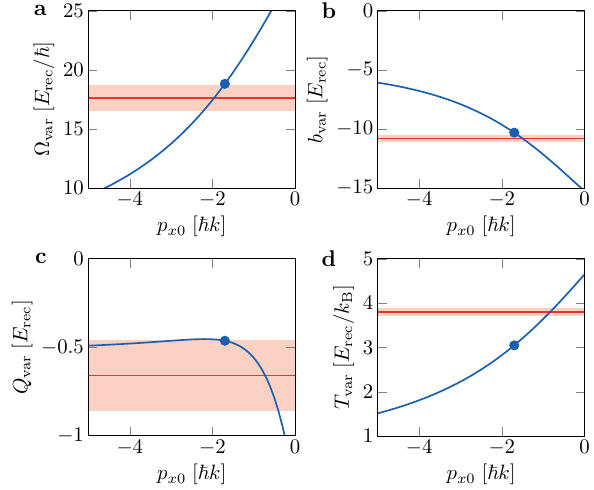}
    \phantomsubfloat{\label{fig:quadratic_approximation:omega}}
    \phantomsubfloat{\label{fig:quadratic_approximation:b}}
    \phantomsubfloat{\label{fig:quadratic_approximation:Q}}
    \phantomsubfloat{\label{fig:quadratic_approximation:T}}
\caption{\textbf{Quadratic approximation of $\KABW$.}
Parameters $\Omegavar$, $\bvar$, $\Qvar$ and $\Tvar$ defining the quadratic Hamiltonian $\KAvar$ approximating $\KABW$ around a given momentum $\pxzero$ (blue lines). The value $\pxzero=-1.6\,\hbar k$ that best accounts for our measurements is shown as blue dots. The red line indicates the experimentally determined optimal parameters, the light red area indicating the error bar.
\label{fig:quadratic_approximation}
}
\end{figure}

\sectionmethods{Quadratic approximation of the BW Hamiltonian}\\[-8mm]
Our variational Hamiltonian $\KAvar$ contains only operators quadratic in $\px$ and $J_z$, while the BW Hamiltonian $\KABW$ include cubic operators as well. To establish a connection between the two, we approximate the cubic operators with quadratic ones, by expanding them to second order in powers of  $\px-p_{x0}$, and $J_z-\mzzero$, where $p_{x0}$ is a chosen momentum and $\mzzero$ is the mean spin projection $\langle J_z\rangle$ for that momentum. Explicitly, we write
\begin{align*}
 J_z^3\simeq&\;\mzzero^3+3\mzzero^2 (J_z-\mzzero)+3\mzzero (J_z-\mzzero)^2,\\
 J_z^2p_x\simeq&\; \mzzero^2\pxzero+2\mzzero\pxzero(J_z-\mzzero)+\mzzero^2(\px-\pxzero)\\
 &+\pxzero(J_z-\mzzero)^2+2\mzzero(\px-\pxzero)(J_z-\mzzero),\\
 J_z p_x^2\simeq&\; \mzzero\pxzero^2+2\mzzero\pxzero(\px-\pxzero)+\pxzero^2(J_x-\mzzero)\\
 &+\mzzero(\px-\pxzero)^2+2\pxzero(\px-\pxzero)(J_z-\mzzero).
\end{align*}
This leads to an approximate version of $\KABW$ $\KABW$ that belongs to the family of variational Hamiltonians studied experimentally. We expect the approximated Hamiltonian to be close to the original BW Hamiltonian for momenta $\px$ close to $\pxzero$.

We show in \fig{fig:quadratic_approximation} the evolution of the variational parameters with $\pxzero$. The choice $\pxzero=-1.6\hbar k$, close to the middle of the momentum interval used for the variational optimization, accounts well for the optimal values determined experimentally. This illustrates the importance of the BW Hamiltonian in our approach.

\sectionmethods{Local inverse temperature $\beta(m)$}
We fit the local inverse temperature profile $\beta(m)$ so that the theoretical deformed Hamiltonian $\{\beta(J_z),H\}$ best fits the optimal variational Hamiltonian $\KAvar$. For this, we  minimize the relative entropy $S(\rhovar||\rho_\beta)$ between the thermal density matrices of the two Hamiltonians (still computed over the momentum interval $-10\hbar k<\px<5\hbar k$).

\begin{figure}
\includegraphics[scale=0.95,trim=3mm 0 0 0]{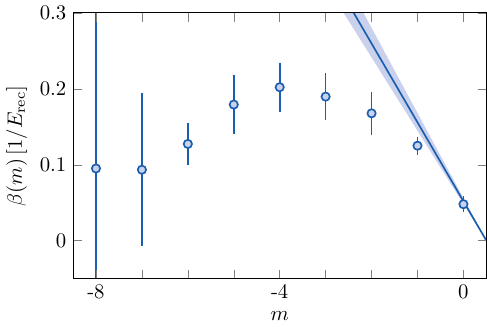}
\caption{\textbf{Local inverse temperature.}
Local inverse temperature $\beta(m)$ fitted to minimize the relative entropy $S(\rhovar||\rho_\beta)$ between thermal states of our experimental optimum variational Hamiltonian and a theoretical deformed Hamiltonian $\{\beta(J_z),H\}$.
Statistical error bars are computed using a bootstrap sampling procedure. The solid line is the linear variation expected for the BW Hamiltonian.
\label{fig:local_beta}
}
\end{figure}

In practice, allowing each individual value of $\beta(m)$ to be a free parameter can result in significant uncertainty in the fit. To address this, we assume a polynomial form for $\beta(m)$ of order 6 in $m$, with the constraint $\beta(0)=-\beta(1)$ required to ensure no coupling between $A$ and $A^c$. The resulting inverse temperature profile $\beta(m)$ is shown in \fig{fig:local_beta}. It agrees well with the expected linear variation from the BW Hamiltonian for magnetic projections near $\mstar$. This further shows the similarity between the optimal variational Hamiltonian and $\KABW$.

\begin{figure}
\includegraphics[scale=0.92,trim=3mm 0 0 0]{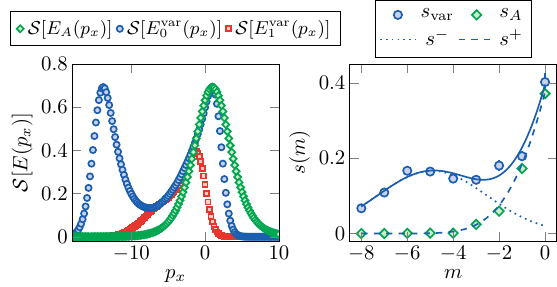}
    \phantomsubfloat{\label{fig:entropy:sp}}
    \phantomsubfloat{\label{fig:entropy:sm}}
    \vspace{-2\baselineskip}
\caption{\textbf{Thermodynamic and entanglement entropies.}
\captionlabel{fig:entropy:sp}
Momentum contributions to the entanglement entropy $\mathcal{S}[E_A(\px)]$ (green diamonds) and to the thermodynamic entropy $\mathcal{S}[E_0^{\text{var}}(\px)]$ (ground band, blue circles) and $\mathcal{S}[E_1^{\text{var}}(\px)]$ (first excited band, red squares).
\captionlabel{fig:entropy:sm} Local entanglement entropy of  the quantum Hall insulator (red squares), compared with the local entropy of a thermal state of $\KAvar$ (blue disks), together with a double-structure fit.
\label{fig:entropy_momentum}
}
\end{figure}

\sectionmethods{Thermodynamic and entanglement entropies}
Our entanglement Hamiltonian implementation allows us to measure the entanglement entropy $S_A$, which maps  to a thermodynamic entropy that is accessible in ultracold atomic gases with well-established protocols \cite{luo_measurement_2007,yefsah_exploring_2011}. We consider a thermal ensemble of fermions in the state  $\rhovar$ as defined in Eq. \eqP{eq:rhovar} in the main text. Its entropy can be expressed in terms of the bandstructure $E_n^{\text{var}}(\px)$ of the adimensional Hamiltonian $\KAvar$, as
\[
 S_{\text{var}}=\sum_{n=0}^\infty\int\frac{\dd p_x}{2\pi}\mathcal{S}[E_n^{\text{var}}(\px)],
\]
where $\mathcal{S}(E)=\mathcal{H}[1/(1+\E^E)]$ represents the contribution to entropy from a state of energy $E$.

We show in \fig{fig:entropy:sp} the variation of $\mathcal{S}[E_n^{\text{var}}(\px)]$ for the ground and first excited bands, as well as the equivalent quantity $\mathcal{S}[E_A(\px)]$ for the pseudo-spectrum $E_A$. In both cases, the entropy is computed from the measured dispersion relations $E_n^{\text{var}}(\px)$ (for $n=0,1$) and $\EA(\px)$. The ground band contribution $\mathcal{S}[E_0^{\text{var}}(\px)]$ exhibits a double structure, with excitations around the Fermi point $\pFm$ (resp. $\pFp$) for $\px<-7\hbar k$ (resp. $\px>7\hbar k$). By summing over excitations around $\pFp$ only, we obtain an entropy per unit length $S_{\text{var}}^{(0+)}\simeq0.65(5)\,k$, which is consistent with the entanglement entropy $S_A\simeq0.66(3)\,k$. In the same momentum interval, the contribution from the first excited band amounts to $S_{\text{var}}^{(1+)}\simeq0.34(2)\,k$. We have evaluated the contribution of higher bands ($n\geq 2$) using the theoretical bandstructure of $\KAvar$. The results indicate that these higher bands do not significantly contribute to the entropy.

While the previous calculation allowed us to distinguish the entropy coming from the Fermi point $\pFp$ from the other Fermi point $\pFm$ and from the first excited band, an actual thermodynamic measurement of the total entropy would not separate these different contributions. This motivates the measurement of the local entropy, which is also experimentally accessible. It can be expressed as (neglecting bands $n\geq2$)
\[
 s_{\text{var}}(m)=\sum_{n=0,1}\int\frac{\dd p_x}{2\pi}\mathcal{S}[E_n^{\text{var}}(\px)]\Pi_n(\px,m),
\]
which involves the  spin projection probabilities $\Pi_n(\px,m)$. While the measurement of the ground-band  probabilities $\Pi_0(\px,m)$ has been presented in the main text, the probabilities $\Pi_1(\px,m)$ are measured from cyclotron excitation dynamics, as discussed in the corresponding section of the Methods.
The local entropy computed from these experimental data, plotted in \fig{fig:entropy:sm}, exhibits a double structure, which we fit by the function
$
s_{\text{var}}(m)=s^{-}(m)+s^{+}(m),
$
where the function $s^{\pm}(m)$ is the entropy density expected for a continuous Hall system, with a linear dispersion relation around the Fermi point $\px^{\pm}$. The area of the fit $s^{+}(m)$ gives a value $S_{\text{var}}^{+}\simeq0.7(1)\,k$, smaller than the sum $S_{\text{var}}^{(0+)} + S_{\text{var}}^{(1+)}=0.95(4)\,k$ introduced earlier. This difference arises because the local entropy from the first excited band is broadly distributed along $m$ and is less accounted for by the fit. This explains why the local entropy fit gives a thermodynamic entropy $S_{\text{var}}^{+}$ very close to the actual entanglement entropy $S_A$.

\end{document}